\documentclass[amstex,aps,showpacs,preprint]{revtex4}%
\usepackage{graphicx}
\usepackage{amsmath}
\usepackage{bm}%
\usepackage{amsfonts}%
\usepackage{amssymb}

\begin{document}

\title{Rydberg wavepackets in terms of hidden-variables: de Broglie-Bohm trajectories}

\author {A.\ Matzkin}
\affiliation{Laboratoire de Spectrom\'{e}trie physique (CNRS
Unit\'{e} 5588), Universit\'{e} Joseph-Fourier Grenoble-I, BP 87,
38402 Saint-Martin, France}

\begin{abstract}
The dynamics of highly excited radial Rydberg wavepackets is
analyzed in terms of de Broglie-Bohm (BB) trajectories. Although
the wavepacket evolves along classical motion, the computed BB
trajectories are markedly different from the classical dynamics:
in particular none of the trajectories initially near the atomic
core reach the outer turning point where the wavepacket localizes
periodically. The reasons for this behavior, that we suggest to be
generic for trajectory-based hidden variable theories, are
discussed.
\end{abstract}

\maketitle

\section{Introduction}

A Rydberg atom is an effective one-electron atom.\ Following laser
excitation, a single electron (called the Rydberg electron) is
excited and periodically returns to the region near the core (made
up of the nucleus and the other tightly bound electrons). When the
laser pulse is short-lived, a wavepacket is formed. Since they
were first observed some 15 years ago \cite{rad1,rad2}, the
production and detection of Rydberg wavepackets have become
increasingly accurate, with developments pointing toward possible
applications in quantum information \cite{bucksbaum00} or
fundamental investigations such as the production of
Schr\"{o}dinger cat states \cite{noel stroud96}. To analyze
Rydberg wavepacket propagation, the main theoretical frameworks
from the early analysis \cite{averbukh-perelman} down to the more
sophisticated models \cite{zoller,robinett} have called for
classical or semiclassical techniques and concepts, usually
encapsulated within the path integral formulation of quantum
mechanics: an initially localized and highly excited wavepacket is
said to evolve\ along classical trajectories, the observed
interferences being due to pieces of the wavepacket that travel
along different classical trajectories.

An alternative interpretation of quantum phenomena hinges on the
existence of particles each following a well-defined space-time
trajectory.\ The de Broglie Bohm (BB) theory is by far the
best-known and most developed of the trajectory-based hidden
variable theories \cite{holland93}. BB trajectories have been
determined for a wide range of quantum systems, including
atom-surface diffraction \cite{sbm01}, quantum billiards
\cite{alcantara00}, kicked rotators \cite{pla298} or even photons
\cite{pla290}. Surprisingly, little has been done on Rydberg atoms
\cite{gocinski99,pla04} and even there only elliptic states
(angular wavepackets) were treated: BB trajectories for angular
wavepackets were seen to closely approximate classical motion.
However radial wavepackets are the most accessible experimentally,
since they only require an optical excitation from the ground
state of the atom. Indeed, most fundamental properties of
wavepackets have essentially been investigated by forming radial,
rather than angular wavepackets.

The object of this work is to analyze the dynamics of Bohmian
particles in a radial wavepacket. To this end, we will sketch in
Sec. 2 a realistic model of laser excitation resulting in the
creation of a radial wavepacket in the hydrogen atom, also valid
for other Rydberg atoms with a nonpenetrating outer electron (i.e.
small quantum defects). We will give a brief overview of Bohmian
dynamics (Sec. 3) and then compute the BB trajectories for the
Rydberg  electron (Sec. 4): we will see that the BB\ trajectories
get quickly uncorrelated with the center of the wavepacket.\ In
particular, none of the BB\ trajectories initially in the core
region reach the outer tuning point, where the wavepacket
localizes after half the classical period. The reasons for this
behavior, some of which will be argued to be relevant to the model
employed and others intrinsic to trajectory based hidden variable
theories, will be discussed in Sec. 5. Our closing comments will
be given in Sec. 6.

\section{Generation and propagation of Rydberg Wavepackets}

We describe in this Section the excitation process leading to the
creation of a wavepacket.\ We will insist on the main physical
ideas (a detailed theoretical account may be found in
\cite{zoller}).\ We assume the atom is initially in its ground
state, that we take to be represented by the hydrogenic ground
state wavefunction $1s$, radially given by a decreasing
exponential with mean value $\bar{r}=1.5$ au and with negligible
probability density for $r\gtrsim5$ au (atomic units will be used
throughout). The exciting laser pulse is described in the
weak-interaction limit by the electric field
$\mathbf{E}=\mathcal{E}(t)e^{-i\omega t}\mathbf{e}$, where
$\omega$ is the laser frequency, $\mathbf{e}$ the polarization
vector and $\mathcal{E}(t)$ the excitation enveloppe. For
definiteness, we shall set $\mathbf{e}$ to be linearly polarized,
$\omega$ to correspond to a transition to the Rydberg state $n=40$
and choose for the excitation profile a Gaussian enveloppe.
Provided the duration $\tau_{P}$ of the Gaussian pulse is
sufficiently short (much shorter than the recurrence time of the
wavepacket), the wavefunction of our system can be written as
\cite{zoller}%
\begin{equation}
\left|  \psi(t)\right\rangle =\chi(t)\left|  1s\right\rangle +\int_{-\infty
}^{t}\left|  \phi(t^{\prime})\right\rangle dt^{\prime}, \label{e2}%
\end{equation}
where the time line has been set as follows. At $t=-\infty$ the atom is in the
ground state; at $t=0$ the maximum of the pulse has reached the atom.
$\chi(t)$ represents the depletion of the initial state that gets excited by
the laser; it stands as a cut-off function, with $\chi(t\ll-\tau_{P}%
/2)=1$.\ Finally the second term on the right handside of Eq. (\ref{e2}) is
the excited wavepacket.\ This term is of course zero for $t\ll-\tau_{P}/2$. On
the other hand, for $t\gg\tau_{P}/2$, i.e. after the pulse excitation, it
takes the form%
\begin{equation}
\int_{-\infty}^{t\gg\tau_{P}/2}\left|  \phi(t^{\prime})\right\rangle
dt^{\prime}=\sum_{n}c_{n}e^{-iE_{n}t}\left|  R_{n}\right\rangle \left|
l=1,m=0\right\rangle , \label{e4}%
\end{equation}
where $\left|  R_{n}\right\rangle $ is the hydrogenic radial wavefunction
(regular Coulomb function) corresponding to principal quantum number $n$ and
$l=1.$The coefficients $c_{n}$, which are related to the Fourier transform of
the excitation enveloppe, take here the Gaussian form%
\begin{equation}
c_{n}=d_{n}\exp\frac{(n-n_{0})^{2}}{2\Delta n^{2}}. \label{e6}%
\end{equation}
The center of the Gaussian distribution $n_{0}$ depends on the
laser frequency $\omega$ (in this work we set $n_{0}=40)$ and the
width $\Delta n$ on the pulse duration $\tau_{P}$. $d_{n}$ is the
normalizing factor, which strictly speaking also depends on the
dipole excitation strength to the excited states $\left|
R_{n}\right\rangle \left| l=1,m=0\right\rangle .$

\begin{figure}[tb]
\includegraphics[height=3.5in,width=2.3in]{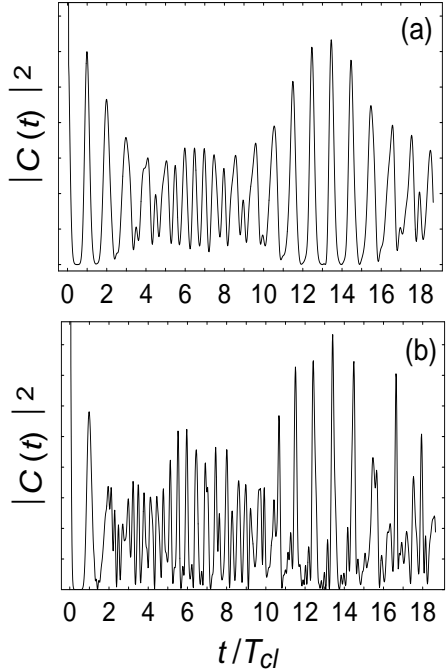}
\caption[]{The autocorrelation function $|C(t)|^2$ (arbitrary
units) for a wavepacket with mean energy $E_0$ is given as a
function of time in units of the classical period of the electron
for $n_0=40$, $T_{cl}\thickapprox 9.7$ ps. (a) corresponds to
$\Delta n=.75$ and (b) to $\Delta n=1.5$, yielding a larger energy
excitation window. \label{fig1}}
\end{figure}

In short, the laser pulse creates a radial wavepacket whose expansion on the
eigenstates is given by a Gaussian distribution. The wavepacket propagates
outward, reaches the outer turning point and comes back toward the core
region. The experimental generation and detection of such wavepackets is very
common \cite{jones98}.\ The autocorrelation function is of particular
interest.\ It is given by the overlap of the propagating wavepacket with the
initially excited one,%
\begin{equation}
C(t)=\left\langle \psi(t=0)\right|  \left.  \psi(t)\right\rangle =\sum
_{n}\left|  c_{n}\right|  ^{2}e^{-iE_{n}t} \label{e8}%
\end{equation}
and is straightforward to compute. Experimentally this quantity is
directly detected by employing a second excitation laser
\cite{jones98}.

Fig.\ 1 shows two examples of autocorrelation functions computed
with the characteristics detailed above.\ The only difference is
the pulse duration $\tau_{P}$ which is shorter in Fig. 1(b),
leading to a wider excitation window $\Delta n$. This is why the
wavepacket spreads faster.\ Still in both Figs. 1 (a) and (b), the
characteristic time scales are visible: at short times the
wavepacket is rather well-localized and the first peaks of $C(t)$
(in particular in Fig. 1(a)) are produced by the return of the
wavepacket to the core at the classical period
\begin{equation}
T_{cl}(E_{0})=\pi  2^{-1/2}(-E_{0})^{3/2},
\end{equation}
where $E_{0}=-1/2n_{0}^{2}$ is the energy of the center of the
wavepacket. The classical periodicity arises from Eq. (\ref{e8})
by employing semiclassical arguments: in a WKB approach, expanding
$E_n$ and using the WKB quantization condition gives the
dependence $t/T_{cl}$ in the exponential
\cite{averbukh-perelman,robinett}. Alternatively, imposing the
stationary phase approximation in the Feynman path integral
expression for propagator leads when $t$ is a multiple of $T_{cl}$
to isolated peaks or to a complex interference pattern, depending
on whether different bits of the wavepacket carried by classical
trajectories having different energies interfere appreciably.
Another feature of the autocorrelation function concerns the
revival phenomenon, particularly visible at $t/T_{cl}> 13$ in Fig.
1(a), due to the fact that the different classical paths
approximately recover the original phase relationships, yielding a
wavefunction nearly identical to the original one. The position of
the wavepacket at different times can also be computed (in
principle it can also be detected experimentally \cite{burgdofer
etal01}).\ This has been done in Fig.\ 2, where the wavefunction
is plotted at different times shortly after excitation. It can be
seen that once a localized wavepacket gets formed it moves in
accordance with classical dynamics. At $t=T_{cl}/2$, more than
99\% of the wavefunction lies beyond $r=1600$ au.
\begin{figure}[tb]
\includegraphics[height=3.8in,width=1.75in]{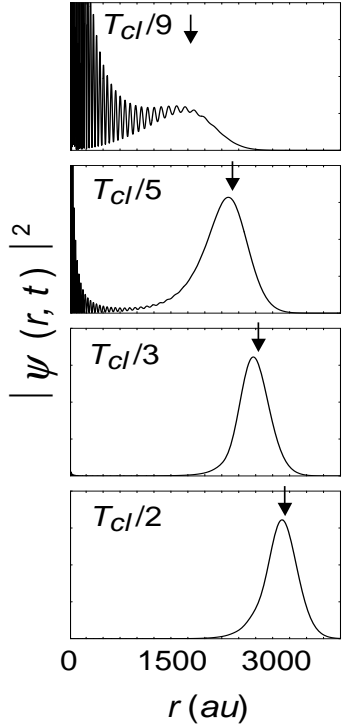}
\caption[]{Evolution of the radial wavepacket corresponding to
case (b) of Fig. 1. Snapshots of the wavefunction are taken at
different times. Each arrow shows the position at the
corresponding time of an electron along a \emph{classical}
trajectory with initial position $r_{cl}=2$ au. The turning point
for the center of the wavepacket lies at
$r_{\mathrm{tp}}\thickapprox 3200$ au.\label{fig2}}
\end{figure}

\section{Dynamics with a single hidden trajectory: De Broglie-Bohm mechanics}

In his original paper Bohm \cite{bohm52} proposed an interpretation by which
an individual quantum system has a ''precise behavior'' in terms of hidden
variables. The standard quantum formalism, that reaches an excellent agreement
with the statistical distribution of measurements would thus be explained by
averaging over a distribution of well defined but hidden trajectories. In the
last decade, BB mechanics has become increasingly popular, and excellent
accounts of the interpretation are available \cite{holland93,bohm hiley}. The
main dynamical equations arise from the polar decomposition of the
wavefunction in configuration space. Put
\begin{equation}
\psi(\mathbf{r},t)=\rho(\mathbf{r},t)\exp(i\sigma(\mathbf{r},t)) \label{e30}%
\end{equation}
where $\rho$ and $\sigma$ are real functions. The Schr\"{o}dinger equation
becomes equivalent to the coupled equations%
\begin{align}
\frac{\partial\sigma}{\partial t}+\frac{(\triangledown\sigma)^{2}}%
{2}+V-\frac{\triangledown^{2}\rho}{2\rho}  &  =0\label{e33}\\
\frac{\partial\rho^{2}}{\partial t}+\triangledown(\rho^{2}\triangledown
\sigma)  &  =0. \label{e34}%
\end{align}
$\rho^{2}$ gives the statistical distribution of the particle (here the
Rydberg electron) whereas the trajectory is obtained by integrating the
equation of motion%
\begin{equation}
\frac{d\mathbf{r}}{dt}=\mathbf{\triangledown}\sigma(\mathbf{r},t) \label{e10}%
\end{equation}
where the initial position of the electron $\mathbf{r}(t=-\infty)$ lies within
the initial distribution $\rho(t=-\infty).$

As should be expected from any theory that accounts for quantum dynamics by
employing a single trajectory, there is a close relationship between the
motion of the putative particle and the net probability current. Eq.
(\ref{e34}) expresses the conservation of the probability flow. For an energy
eigenstate of a Rydberg atom, there is no net probability flow in the radial
coordinate and according to Eqs. (\ref{e30})-(\ref{e34}) the electron has no
radial motion. This is dynamically explained by the last term on the right
handside of Eq. (\ref{e33}),
\begin{equation}
Q(\mathbf{r},t)=-\frac{\triangledown^{2}\rho}{2\rho}\label{e70}%
\end{equation}
known as the quantum potential, that cancels the effects of the
classical potential $V$.

It is thus interesting to compute BB\ trajectories for time
dependent wavepackets and to give an interpretation of an
observable quantity such as the autocorrelation function in terms
of the statistical distribution of hidden but well-defined
electron trajectories. This is readily done by casting the Rydberg
wavepackets obtained in Sec.\ 2 in the form of Eq. (\ref{e30}).
Since we have set $m=0$, $\sigma(\mathbf{r},t)$ has no
contribution from the angular coordinates and there is no angular
motion: once an initial position has been set for the particle,
the electron incurs only a radial motion. Eq. (\ref{e10}) is thus
simply one dimensional. However it is still demanding to solve it
because the equation of motion is extremely sensitive to the
position of the nodes of the wavefunction. Not only do highly
excited eigenstates, from which the wavepacket is built, have a
great number of nodes, but the 'quasi-nodes' of the wavepacket can
move very quickly in particular during the application of the
laser pulse.

\section{Results}

We have numerically integrated Eq. (\ref{e10}) for the two Rydberg
wavepackets (a) and (b) of Sec. 2. The resulting de Broglie-Bohm
trajectories are shown in Fig. 3. We have chosen as the initial
condition $r_{BB}(t=-\infty)=2$ au. Recall that the initial state
$\left| 1s\right\rangle $ is compactly localized around the core
($0<r\lesssim 10,$ $\bar{r}=1.5$ au) and that BB\ trajectories
cannot cross: if initially $r_{1}(-\infty)>r_{2}(-\infty)$ then
$r_{1}(t)>r_{2}(t)$ for all $t$. During the application of the
laser pulse $r_{BB}(t)$ varies erratically back and forth around
the initial position with a small amplitude. The particle then
starts to move outward appreciably (for $t\gtrsim T_{cl}/10$). At
$t\approx T_{cl}/2$, $r_{BB}$ reaches a maximum value
$r_{BB}^{\max}$ (much larger for the case (b) wavepacket) and
turns back toward the core. Indeed at that time the wavepacket has
reached its maximum distance from the core and is localized around
the classical outer turning point $r_{\mathrm{tp}}=3200$ au (see
Fig. 2). Note however that $r_{BB}^{\max}\ll r_{\mathrm{tp}}$: the
BB trajectory never reaches the outer
turning point. This remains true whatever the initial position $r_{BB}%
(t=-\infty)$ of the electron (but see Discussion below). At
$t=T_{cl}$ part of the wavepacket has returned to the core,
producing a peak in the autocorrelation function (Fig.\ 1); the
BB\ trajectory has also returned to the core region,
$r_{BB}(T_{cl})$ being slightly larger than $r_{BB}(-\infty)$.

For longer times, the trajectories obtained for the wavepackets
(a) and (b) are very different reflecting the difference in the
quantum potential arising from the difference in the interference
pattern (remember the only difference between the wavepackets (a)
and (b) is the excitation window $\Delta n$). This is illustrated
in the insets in Figs. 3(a) and 3(b) for the interval
$2<t/T_{cl}<3$: in case (b) $r_{BB}(t)$ oscillates back and forth
several times whereas in case (a) the trajectory makes a single
round trip from the core, reflecting the classical periodicity.
Note also that the revivals observed in the autocorrelation
function (Fig.\ 1) are not neatly visible in the behavior of the
BB trajectories.\ The reason is that revivals occur when the
wavepacket \emph{approximately} regain its initial shape, but then
the quantum potential can be quite different.

\begin{figure}[tb]
\includegraphics[height=4in,width=2.9in]{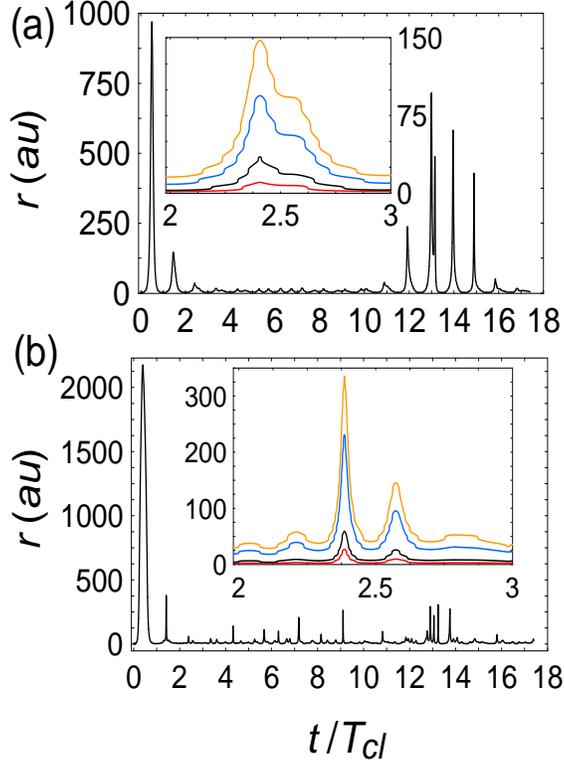}
\caption[]{Trajectories $r_{BB}(t)$ computed according to de
Broglie-Bohm dynamics. (a) shows the trajectory with initial
position $r_{BB}=2$ au as a function of time for the wavepacket
(a) (see Fig. 1). The inset zooms on the interval $2<t/T_{cl}<3$
and in addition shows in grey (or color) BB trajectories evolved
from different initial positions: from top to bottom the initial
positions corresponding to the trajectories in the inset are 10
(yellow), 6 (blue), 2 (black) and 1 (red) au. (b): same as (a) but
for the case (b) wavepacket. \label{fig3}}
\end{figure}

Fig. 4\ shows the situation for short times, i.e. when the
wavepacket is still fairly well localized, for the case (b)
wavepacket. The classical and Bohmian trajectories are plotted
along with the average position of the wavepacket (expectation
value of the position operator). The situation portrayed in Fig.\
4\ is an application of Ehrenfest's theorem: before the break time
the wavepacket evolves along a classical trajectory, whereas as
can be easily shown (see Sec.\ 3.8 of \cite{holland93}) BB
trajectories generically never move with the wavepacket.

\begin{figure}[b]
\includegraphics[height=1.9in,width=2.5in]{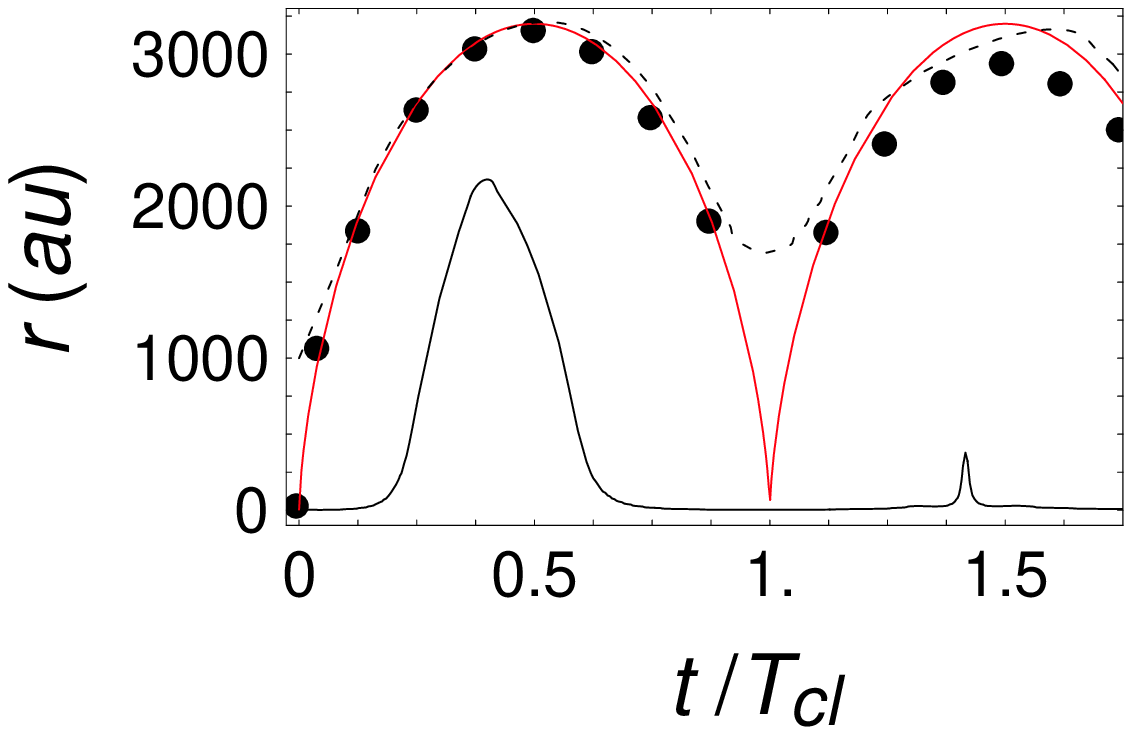}
\caption[]{Short-time evolution of classical and Bohmian
trajectories and average position of the wavepacket for the case
(b) wavepacket. The black solid line is the BB trajectory with
initial value $r_{BB}=2$ au already shown in Fig.\ 3(b), and the
gray (red) line is the classical trajectory with the same initial
position. The black dots represent the average position of the
wavepacket. The dashed line is the BB trajectory that reaches the
outer turning point region along with the wavepacket. Its
'initial' (at $t=0$) position is $r\simeq1000$ au.\ \label{fig4}}
\end{figure}

\section{Discussion}

We have seen that the BB trajectories computed above have a striking property:
no trajectory lying initially in the core region ever reaches the region near
the outer turning point where the wavepacket periodically gets localized. We
now discuss to what extent this is a feature specific to the present model or
can be said to be an intrinsic aspect of the de Broglie-Bohm dynamics, or more
generally of trajectory-based hidden variable theories.

The model summarized in Sec.\ 2 above is the most widely employed
model describing Rydberg excitation by a tailored laser pulse.\
Crudely speaking, the excited wavepacket is progressively turned
on: it is given by Eq. (\ref{e4}) multiplied by some function
$\zeta(t)$ whose exact form depends on $\mathcal{E}(t),$ but that
obeys $\zeta(t)=0$ for $t<-\tau_{P}/2$ and goes to $\zeta(t)=1$
for $t>-\tau_{P}/2$. Now the wavefunction given by Eq. (\ref{e4})
at $t=0$ is far from being perfectly localized near the origin and
has a small but non-negligible probability density even at large
$r$. This means that at say $r=1000$ au, although initially
($t<-\tau_{P}/2$) there is zero probability density, a small
probability density appears as $\zeta(t)$ increases, which in a
sense 'pops out' of nowhere. This feature arises from the model
employed to describe the wavepacket. In BB\ terms, this gives rise
to trajectories such as the dashed line of Fig.\ 4, which obeys $r_{BB}%
(t=0)=1000\;$au.\ For such a trajectory it is numerically
impossible to integrate the equation of motion backwards in time
for $t\ll0$ since the probability vanishes as $t \rightarrow
-\infty$ (but clearly there cannot be any significant motion from
the position at $t=0$ given that the nodes on each side of that
initial position do not move as $t\rightarrow-\tau_{P}/2$ although
the probability amplitude does decrease to $0$). Note that within
the de Broglie-Bohm interpretation these type of trajectories need
to be taken into account for consistency, because it is only by
including them that the quantum mechanical expectation values can
be matched to the statistical average over the ensemble of
trajectories, as required by the theory.

However, the fact the hidden-variable BB trajectory does not
follow the apparent dynamics of the wavepacket, which when
localized moves in accordance with the classical trajectory, is
unrelated to any specific model of wavepacket propagation. Let us
take an eigenstate of the Hamiltonian: it is a standing-wave,
which except for a global phase-factor (none in the present
example because $m=0$) is a real function. According to the de
Broglie-Bohm theory, there is no motion in that situation, the
particle being at rest by Eq. (\ref{e10}). This is generically
true of any theory positing a single trajectory dynamics depending
on the net probability flow, which vanishes for stationary states.
On the other hand, according to the path integral formalism, the
standing wave arises from the interference of two waves travelling
in the opposite directions; from the semiclassical approximation
to the propagator, the waves follow the sole classical periodic
orbit of the system, but interfere. It is clear that positing a
single trajectory in space-time to account for quantum phenomena
proscribes the possibility for such a trajectory to follow
classical motion if interference effects are important.

The last remark also holds for an arbitrary wavepacket. Although
the average position of the wavepacket follows the classical
trajectory very quickly (Fig. 4), the wavepacket only starts to
get localized for $t\gtrsim T_{cl}/4$, as illustrated in Fig.\ 2.
At short times, when the wavepacket is formed, interference
effects are very important. Locally the net probability flow may
be insignificant, although the probability density is large.
According to BB theory, the electron can then hardly move, because
it is 'trapped' by the nodes arising from the interference terms.
By the time the BB\ particle moves away, the bulk of the
wavepacket is already far ahead.\ Indeed, the net probability flow
is delayed relative to the localized part of the wavepacket. At
subsequent times the localized part of the wavepacket accounts for
only a fraction of the total probability density and its motion is
thus uncorrelated with the maximal probability flow.\ For example
in the autocorrelation function shown in Fig.\ 1(a), the classical
periodicity is clearly visible at short times, though only a small
fraction of the total wavefunction returns to the core.\ If one
postulates that the fundamental dynamics relies on the existence
of a single hidden trajectory, whose dynamical law is directly
related to the net probability flow, it is not possible to account
for the classical periodicity (that can be experimentally
observed) by invoking a correspondence with classical dynamics.
Another illustration is afforded by the insets in Fig. 3. The BB
trajectories in 3(a) and 3(b) are markedly different due to the
local space-time structure of the probability flow. In contrast,
the usually employed semiclassical framework ascribes the same
underlying classical dynamics to both wavepackets, the only
difference being the dispersion rate.

\section{Conclusion}
We have investigated highly excited radial Rydberg wavepackets in
terms of hidden-variables trajectories obtained within the de
Broglie-Bohm framework. The Bohmian picture of Rydberg atoms is
internally consistent and compatible with the experimentally
observed autocorrelation function. However contrarily to angular
Rydberg wavepackets \cite{gocinski99,pla04}, the BB trajectories
have little in common with classical motion. We have suggested
that this is a generic feature of hidden variable theories
accounting for quantum phenomena by the existence of a localized
particle following a single trajectory. On the other hand we have
also seen that the most commonly employed framework to interpret
excited wavepacket phenomena hinges on semiclassical arguments, by
which the dynamics is classical but the different paths interfere
as the wavepacket spreads.

\end{document}